\documentclass[aps,pra,a4paper,preprint]{revtex4}

\usepackage{amssymb}
\usepackage{amsmath}
\usepackage{bm}
\usepackage{natbib}
\usepackage{graphics,graphicx}
\usepackage{epsfig}
\usepackage{latexsym}
\usepackage{epic,eepic,graphicx,amssymb,amsmath,indentfirst}
\usepackage{bm}
\usepackage{graphicx}
\usepackage[utf8]{inputenc}
\usepackage[T1]{fontenc}
\usepackage{natbib}

\usepackage{dsfont}

\usepackage{graphicx}
\usepackage{float}
\usepackage[utf8]{inputenc}
\usepackage[T1]{fontenc}
\usepackage{amsmath, amssymb, amsfonts}
\usepackage{braket} 
\usepackage{geometry}
\usepackage{yfonts}
\DeclareMathOperator{\sgn}{sgn}

\usepackage{multirow}

\usepackage{color}
\usepackage[colorlinks={true}]{hyperref}
\hypersetup{citecolor={blue}, filecolor={blue}, linkcolor={blue}, urlcolor={blue}}

\begin{document}

\title{\bf Reverse engineering of single-qubit quantum gates}


\author{Gustavo Fernandes da Costa}
\author{Leonardo K. Castelano}
\author{Emanuel Fernandes de Lima}
\email{eflima@ufscar.br}
\affiliation{Departamento de F\'isica, Universidade Federal de S\~ao Carlos (UFSCar)\\ S\~ao Carlos, SP 13565-905, Brazil}
\date{\today}

\begin{abstract}
In this work, we address the problem of designing single-qubit quantum gates by means of a linearly-polarized field. We show that any desired one-qubit gate corresponding to a special unitary matrix can be generated by a modulated sinusoidal field. The only approximation involved is the rotating-wave-approximation. The formula for the control field is obtained by inverting the equation of motion for the evolution operator and imposing the conditions for the desired gate. We give a simple procedure to obtain closed analytical formulas for the fields in terms of \textit{a priori} chosen dynamical functions. Additionally, these dynamical functions can depend on tunable free parameters intentionally introduced to meet a desired performance criteria.
\end{abstract}
\maketitle

\section{Introduction}

Building high fidelity quantum gates is a fundamental task for quantum computing \cite{PhysRevApplied.21.064048,George2025,h4wk-v31j}. These logical gates are generated in practice by a set of electromagnetic control pulses. In the case of single-qubit gates, which are building blocks of quantum circuits, constructing arbitrary gates with a sequence of pulses is in principle straightforward \cite{Nielsen2000}. Indeed, any rotation in the Bloch sphere can be performed with a three-pulse sequence, combining for instance, $Z$ and $Y$ rotations \cite{RevModPhys.79.135,HAFFNER2008155,Saffman_2016,PhysRevLett.100.247001}. But using a sequence of pulses have some drawbacks, such as the synchronization of the pulses and the consequent increasing of the hardware complexity. Additionally, a sequence of pulses can increase the gate operation time, as compared to a single pulse, and also decrease the overall fidelity, since it is the product of the fidelity of each step \cite{PhysRevB.85.241304}.
 
Single linearly-polarized field with modulated frequency and amplitude can be used to circumvent these issues \cite{PhysRevA.96.012326,PhysRevA.96.022309,PhysRevLett.111.050404}. Quantum optimal control provides a framework to find a control pulse to generate any single-qubit gate \cite{1367-2630-12-7-075008,Ansel_2024}. Within this approach, one has often to rely on numerical optimization techniques, such as GRAPE, CRAB or Krotov \cite{10.21468/SciPostPhys.7.6.080,Fernandes_2023,PhysRevA.84.022326,Muller_2022,10.1098/rsta.2011.0361,KHANEJA2005296}. Unfortunately, such numerically obtained controls are often complicate and not suitable for experimental realization \cite{Riaz2019,PhysRevA.90.012318}. Thus, it is advantageous to have analytical formulas for the control pulses that generate the quantum gates, both for practical experimental application as well as for fundamental understanding \cite{PhysRevA.96.042339}.    

Alternatively to the optimal control approach, inverse-engineering and related methods allow to obtain the control field in terms of prescribed dynamical functions \cite{PhysRevA.108.033106,PhysRevA.98.043429,PhysRevA.47.4593,PhysRevLett.125.250403,ws7n-3z78}. The central idea is to determine the controls directly from the dynamical equations, which are inverted in terms of the controls. The controls are then found by imposing the desired dynamics through a set of dynamical functions, e.g., observables, quantum amplitudes, etc. This prescription has been successfully applied to obtain analytical control pulses for two-level systems under the rotating-wave-approximation (RWA), with and without dissipation \cite{articleGolubev,PhysRevA.110.022201,x2g9-wpk6,articleAndras,PhysRevA.100.012103,PhysRevA.101.023822}. However, these later works have been focused in state preparation, that is, reaching a desire target state from a given initial condition.

In this paper, we extend previous works by obtaining a closed formula for the control pulses to produce any desired single-qubit quantum gate within the RWA. The general formula is given by a sinusoidal pulse with modulated phase and amplitude. These phase and amplitude are given in terms of \textit{a priori} chosen dynamical function, which must satisfy few constraints. As we show, the freedom to chose the dynamical function can be leverage to minimize a given cost functional.  

\section{Design of the control field}
We consider a two-level system subjected to a linear polarized field $E(t)$ along the $x$-direction, 
\begin{equation}\label{Ham}
    H(t) = H_0 + H_1(t) = -\frac{\omega_0}{2} \sigma_z -\mu E(t) \sigma_x,
\end{equation}
where $\omega_0$ is the system transition frequency and $\mu$ is the dipole coupling with the external field.

Our goal is to design an arbitrary quantum gate $U_{\rm gate}$, which plays the role of evolution operators $U(t_f,0)=U_{\rm gate}$, by means of the control field $E(t)$ acting from $t=0$ to a final time $t=t_f$. This control pulse which generates the gate is obtained from the inversion of the dynamical equations, as explained in the following.

Writing the control pulse as,

\begin{equation}
    E(t) = \varepsilon(t) e^{-i\omega_p t} + \varepsilon^*(t) e^{i\omega_p t},
    \label{eq:E(t)}
\end{equation}
where $\omega_p$ is the pulse carrier frequency. The Hamiltonian in the Dirac interaction picture upon applying the rotating-wave-approximation (RWA) is given by,
\begin{equation} \label{HamRWA}
    \tilde{H}(t) = \begin{pmatrix} 0 & -\mu \varepsilon^*(t)e^{i(\omega_p-\omega_0)t} \\ -\mu \varepsilon(t)e^{i(\omega_0-\omega_p)t} & 0 \end{pmatrix}
\end{equation}

Although an evolution operator is a unitary matrix, here, we restrict ourselves to the set of special unitary matrices, since all single qubit unitary gates may be written as the product of an element of $SU(2)$  with a global phase factor \cite{Nielsen2000}. Thus, we can write a general evolution operator from time $t=0$ to time $t$ belonging to SU(2) as,

\begin{equation}
    \tilde{U}(t,0) = \begin{pmatrix} A(t) & B(t) \\ -B^*(t) & A^*(t) \end{pmatrix},
    \label{eq:U_matrix}
\end{equation}
where the complex functions $A(t)$ and $B(t)$ are such that $|A(t)|^2+|B(t)|^2=1$.

This evolution operator satisfies the Schrodinger equation (atomic units used throughout),

\begin{equation}\label{Schroeq}
    i\frac{d}{dt} \tilde{U}(t,0) = \tilde{H}(t) \tilde{U}(t,0),
\end{equation}
with initial condition $U(0,0)=\mathds{1}$, where $\mathds{1}$ is the unity matrix.

Substituting Eq.~\eqref{eq:U_matrix} into Schrodinger equation \eqref{Schroeq}, we obtain two independent dynamical equations for the coefficients $A(t)$ and $B(t)$,

\begin{align}
    i \dot{A}(t) &= \mu \varepsilon^*(t) B^*(t) e^{i(\omega_p-\omega_0)t} \label{eqa} \\
    i \dot{B}(t) &= -\mu \varepsilon^*(t) A^*(t) e^{i(\omega_p-\omega_0)t} \label{eqb}.
\end{align}

By dividing one of the above equations by the other, we obtain a second relation among $A(t)$ and $B(t)$,

\begin{equation}
     \dot{A}(t)A^*(t)+\dot{B}(t)B^*(t) = 0.
    \label{eq:A_dot}
\end{equation}

In order to express the field in terms of the functions $A(t)$ and $B(t)$, we multiply Eq.\eqref{eqa} by $B(t)$ and Eq.\eqref{eqb} by $A(t)$ and sum the two equations yielding,
\begin{equation} \label{eq:epsilon_conj}
    \varepsilon^*(t) = -\frac{i}{\mu} \left( A(t)\dot{B}(t) - B(t)\dot{A}(t) \right) e^{-i(\omega_p-\omega_0 )t }.
\end{equation}

From relation \eqref{eq:A_dot} and the special unitary condition, the term in parentheses can rewritten as,
 
\begin{equation}
    A(t)\dot{B}(t) - B(t)\dot{A}(t) =  A(t)\dot{B}(t) - B(t)\left( - \frac{\dot{B}(t)B^*(t)}{A^*(t)} \right) = \frac{\dot{B}(t)}{A^*(t)}.
\end{equation}

Thus, Eq.~\eqref{eq:epsilon_conj} can be expressed as,
\begin{equation}
    \varepsilon^*(t) = -\frac{i}{\mu} \left( \frac{\dot{B}(t)}{A^*(t)} \right)  e^{-i(\omega_p-\omega_0 )t }.
    \label{e_conj_emanuel}
\end{equation}

It is convenient to write the field in terms of real functions, so we use the polar representation,

\begin{equation}
    A(t) = a(t)e^{i\phi_a(t)} \quad {\rm and} \quad B(t) = b(t)e^{i\phi_b(t)},
\end{equation}
with $a(t)$, $b(t)$, $\phi_a(t)$, $\phi_b(t)$ being real functions.

These functions are of course not independent: the special unitary condition yields $a(t)^2+b(t)^2=1$, that can be combined with Eq.\eqref{eq:A_dot} to obtain,

\begin{equation}
    \dot{\phi_b}(t) = \frac{a(t)^2}{a(t)^2 -1}\dot{\phi_a}(t), \label{eq:cond_phi}
\end{equation}
which can be integrated yielding

\begin{equation}
    \phi_b(t)=\int \frac{a(t)^2\dot{\phi}_a(t)}{a(t)^2-1}dt+C_b, \label{eq:intphi}
\end{equation}
with $C_b$ being an arbitrary integration constant. Thus, specifying $a(t)$ and $\phi_a(t)$ sets the function $b(t)$ and $\phi_b(t)$ up to a constant.

Expressing $\varepsilon^*(t)$ in Eq. \eqref{e_conj_emanuel} in terms of the real functions and substituting the result in \eqref{eq:E(t)}, we can write the external field as,
\begin{equation}
    E(t) = -\frac{2}{\mu}{\rm Im}\left\{ \left[ \frac{\left(\dot{b}(t) - i\dot{\phi_b}(t)b(t)\right)} {a(t)}  \right] e^{-i(\phi_a(t) + \phi_b(t))}e^{-i\omega_0 t }\right\}
    \label{eq:E(t)_final}
\end{equation}

In order to simplify the above expression, the complex function $\dot{b}(t) - i\dot{\phi_b}(t)b(t)$ can be written in polar form. However, care must be taken since neither the sign of $\dot{b}(t)$ or $\dot{\phi_b}$ are known a priori. To circumvent this situation, we can write

\begin{align}
     \dot{b}(t) - i\dot{\phi_b}(t)b(t) & =\sgn\{\dot{b}(t)\}\left[\vert\dot{b}(t)\vert-i\dot{\phi}_b(t)b(t)\sgn\{\dot{b}(t)\}\right]= \nonumber \\ 
     & =\sgn\{\dot{b}(t)\}\sqrt{\dot{b}(t)^2+\dot{\phi}_b^2(t)b(t)^2}{\rm e}^{i\lambda(t)},   
\end{align}
where $\sgn$ is the sign function and,

\begin{equation}
    \lambda(t)=\arctan\left\{\frac{\dot{\phi}_b(t)b(t)}{\dot{b}(t)}\right\},
\end{equation}
with $-\pi/2\leq\lambda(t)\leq\pi/2$.

Thus, the control pulse in \eqref{eq:E(t)_final} can be written as,

\begin{equation}
    E(t)=F(t)\sin\left[\omega_0 t+\phi(t)+\lambda(t)\right],\label{eq:field_sin}
\end{equation}
with the phase $\phi$ being the sum of the phases, $\phi(t)=\phi_a(t)+\phi_b(t)$, while the time-dependent amplitude is,

\begin{equation}
    F(t)=\frac{2}{\mu}\sgn\{\dot{b}(t)\}\frac{\sqrt{\dot{b}(t)^2+\dot{\phi}_b^2(t)b(t)^2}}{a(t)}.\label{eq:ampl}
\end{equation}
Note that since the above formula depends on $\dot{b}(t)=-a(t)\dot{a}(t)/b(t)$, to avoid singularities, the function $a(t)$ is such that $\dot{a}(t)/b(t)$ is finite whenever $a(t)=1$, i.e., for all times when $b(t)=0$, also $\dot{a}(t)=0$, such that their ratio is finite. Also, the ratios $\dot{b}(t)/a(t)$ and $\dot{\phi}_b(t)/a(t)$ are finite when $a(t)=0$. One should have these constraints in mind when designing the quantum gate.




We are now in a position to describe the method to design the control pulse aimed at producing the desired quantum gate. Assume that the desired $SU(2)$ target gate is given by,

\begin{equation}
    U_{\rm gate} = \begin{pmatrix} a_g{\rm e}^{i\phi_{ga}} & b_g{\rm e}^{i\phi_{gb}} \\ -b_g{\rm e}^{-i\phi_{gb}} & a_g{\rm e}^{-i\phi_{ga}} \end{pmatrix},
    \label{eq:U_gate}
\end{equation}
with $b_g=\sqrt{1-a_g^2}$.

The initial condition for the evolution operator implies that $a(0)=1$, $\phi_a(0)=0$ and $b(0)=0$, while the initial phase $\phi_b(0)$ is undefined. The control pulse should generate the dynamics such that $\tilde{U}(t_f,0)=U_{\rm gate}$, which implies that $a(t_f)=a_g$, $b(t_f)=b_g$, $\phi_a(t_f)=\phi_{ga}$ and $\phi_b(t_f)=\phi_{gb}$. Thus, the control pulse can be build by choosing a function for a(t) satisfying the boundary conditions $a(0)=1$ and $a(t_f)=a_{g}$. We should also define a second function $\phi_a(t)$ meeting the conditions $\phi_a(0)=0$ and $\phi_a(t_f)=\phi_{ga}$. Alternatively, we can allow the function $a(t)$ to assume negative values if the final phase is displaced by $\pi$. In this case, the final time-conditions to be satisfied are $a(t_f)=-a_g$ and $\phi_a(t_f)=\phi_{ga}\pm \pi$.

Once $a(t)$ and $\phi_a(t)$ are defined, the function $\phi_b(t)$ can be obtained by solving the integral \eqref{eq:intphi}. Note that since we have the freedom to set the arbitrary constant of integration, this constant can be adjusted to meet the condition $\phi_b(t_f)=\phi_{gb}$. Therefore, defining the two functions $a(t)$ and $\phi_a(t)$, solving the integral and adjusting the integration constant provides a general framework to produce a given quantum gate belonging to $SU(2)$ by the field given in \eqref{eq:field_sin}.

Finally, we observe that for some gates the value of either $a_g$ or $b_g$ can be zero and in these cases the corresponding phases $\phi_{ga}$ and $\phi_{gb}$ are undefined. As we will see, in these situations we will have a free parameter that can be chosen at will to generate the gate. 

\section{Cosine solution}

The integral \eqref{eq:intphi} can be calculated analytically for appropriate choices of the functions $a(t)$ and $\phi_a(t)$. Here we illustrate a simple choice for the functions that can be used to build any quantum gate. Consider the following particular functions, 

\begin{equation}
    a(t)=\cos\left(\frac{\theta  t}{2t_f}\right),\label{eq:atp}
\end{equation}
which implies that,
\begin{equation}
    b(t)=\sin\left(\frac{\theta  t}{2t_f}\right),\label{eq:btp}
\end{equation}
where $\theta$ sets frequency of oscillation of these functions and should be adjusted to meet the final time condition $a(t_f)=a_g$. Note that the conditions $a(0)=1$, $b(0)=0$ and $a(t)^2+b(t)^2=1$ are satisfied. For this $a(t)$, Eq.\eqref{eq:cond_phi} indicates a possible choice for $\dot{\phi}_a(t)$ as being,

\begin{equation}
    \dot{\phi_a}(t) = \chi \sin\left(\frac{\theta  t}{2t_f}\right)^2,
\end{equation}
with $\chi$ being a constant. We can integrate the above expression yielding,

\begin{equation}
    \phi_a(t) = \frac{\chi}{2} \left[ t - \frac{t_f}{\theta} \sin \left(\frac{\theta t}{t_f} \right) \right],\label{eq:phiatp}
\end{equation}
where we have set the integration constant to zero in order to fulfill $\phi_a(0)=0$. The constant $\chi$ is obtained from the final condition of the phase $\phi_a(t_f)=\phi_{ga}$,
\begin{equation}
    \chi = \frac{2\phi_{ga}}{t_f \left[ 1 - \frac{\sin(\theta)}{\theta} \right]}.
\end{equation}
Now, from Eq.\eqref{eq:cond_phi}, we have,
\begin{equation}
    \dot{\phi_b}(t) = - \chi \cos\left(\frac{\theta t}{2t_f}\right)^2,
\end{equation}
whose integration leads to
\begin{equation}
    \phi_b(t) = -\frac{\chi}{2} \left[ t + \frac{t_f}{\theta} \sin \left(\frac{\theta t}{t_f} \right) \right] + \xi,\label{eq:phibtp}
\end{equation}
where the constant $\xi$ is determined by the final time condition $\phi_b(t_f)=\phi_{gb}$,
\begin{equation}
    \xi = \phi_{gb} + \frac{\chi t_f}{2} \left[ 1 + \frac{\sin(\theta)}{\theta} \right].
\end{equation}
The sum of the phases reduces to,

\begin{equation}\label{eq:phi_cos}
    \phi(t)=-\frac{\chi t_f}{\theta}\sin\left(\frac{\theta t}{t_f}\right)+\xi,
\end{equation}
and the phase $\lambda(t)$ to,

\begin{equation}\label{eq:lamb_cos}
\lambda(t)=\arctan\left\{-\frac{\chi t_f}{\theta}\sin\left(\frac{\theta t}{t_f}\right)\right\},
\end{equation}
while the field amplitude is given by,

\begin{equation}\label{eq:F_cos}
    F(t)=\frac{\theta}{\mu t_f}\sqrt{1+\left(\frac{\chi t_f}{\theta}\right)^2\sin^2\left(\frac{\theta t}{t_f}\right)}
\end{equation}

Therefore, Equations \eqref{eq:F_cos}, \eqref{eq:lamb_cos} and \eqref{eq:phi_cos} provide the amplitude and time-dependent frequency for the control pulse that generates the desired quantum gate. It is also instructive to calculate the time-dependent detuning given by $\Delta(t)=\dot{\phi}(t)+\dot{\lambda}(t)$,

\begin{equation}
    \Delta(t)=-\chi\cos\left(\frac{\theta t}{t_f}\right)\left[1+\frac{1}{1+\frac{\chi^2t_f^2}{\theta^2}\sin^2\left(\frac{\theta t}{t_f}\right)}\right]
\end{equation}

In the following we construct the solutions for some particular gates.

\subsection{ $i\sigma_x$ gate}

Consider the gate $i\sigma_x$,
\begin{equation}
    U_{\rm gate} = 
    \begin{pmatrix} 0 & i \\ i & 0 \end{pmatrix} = i\sigma_x,
\end{equation}
which implies that $a_g=0$, $\phi_{ga}$ is undefined, $b_g=1$ and $\phi_{gb}=\pi/2$. Thus, we have the final time conditions to $a(t_f)=0$, $b(t_f)=1$, which are satisfied by setting $\theta=\pi$. To meet 
$\phi_b(t_f)=\pi/2$, we must have
\begin{equation}
    \xi=\frac{\chi t_f}{2}+\frac{\pi}{2}=\phi_{ga}+\frac{\pi}{2}.
\end{equation}
Notice that we have the freedom to chose the value of $\chi$ or, equivalently, of $\phi_{ga}$. The simplest choice is $\chi=\phi_{ga}0$, which implies that $\xi=\pi/2$ and also that both phases are constant $\phi_a=0$ and $\phi_b=\pi/2$ (also $\lambda=0$). This choice produces a field with constant amplitude and fixed frequency,
\begin{equation}
    E(t)=\frac{\pi}{\mu t_f}\sin\left(\omega_0 t+\pi/2\right),\label{eq:field_sigma_x}
\end{equation}
which is the known rectangular-shape $\pi$-pulse.

For $\chi\neq0$, the control field has a time-varying amplitude,
\begin{equation}
    F(t)=\frac{\pi}{\mu t_f}\sqrt{1+\left(\frac{2\phi_{ga}}{\pi}\right)^2\sin^2\left(\frac{\pi t}{t_f}\right)}\label{eq:Ftsigma_x}
\end{equation}
and time-varying phases,

\begin{equation}
    \lambda(t)=\arctan\left\{-\frac{2\phi_{ga}}{\pi}\sin\left(\frac{\pi t}{t_f}\right)\right\},\label{eq:lambdasigamax}
\end{equation}

\begin{equation}
    \phi(t) = -\frac{2\phi_{ga}}{\pi} \sin \left(\frac{\pi t}{t_f} \right) +\phi_{ga}+ \frac{\pi}{2}.\label{eq:phi_sigmax}
\end{equation}
Therefore, we have a family of control fields which depends on the parameter $\phi_{ga}$ ( or $\chi$) producing the $i\sigma_x$ gate, the $\chi=0$ member being the rectangular-shape $\pi$-pulse.

We note that the fields to produce the $i\sigma_y$ gate are very similar to the ones that produce the $i\sigma_x$ gate, same amplitude $F(t)$ and phase $\lambda(t)$, the only difference being a constant phase of $-\pi/2$ added to the phase $\phi(t)$ in Eq.\eqref{eq:phi_sigmax}.  






\subsection{Phase gates}

Consider the gate ${\rm e}^{-i\varphi/2}P(\varphi)$ corresponding to an increase of $\varphi$ in the phase,
\begin{equation}
    U_{\rm gate} = 
    \begin{pmatrix} {\rm e}^{-i\varphi/2} & 0 \\ 0 & {\rm e}^{i\varphi/2} \end{pmatrix} = {\rm e}^{-i\varphi/2}P(\varphi).
\end{equation}
We choose $a_g=-1$ and $\phi_{ga}=-\pi-\varphi/2$, while $b_g=0$ and $\phi_{gb}$ is undefined. As noted previously, the switched sign of $a_g$ is compensated by displacing the phase $\phi_{ga}$ by $-\pi$. Setting $\theta=2\pi$, we obtain $\chi t_f=2(\pi-\varphi/2)$ and

\begin{equation}
    \phi(t)=\left(1+\frac{\varphi}{2\pi}\right)\sin\left(\frac{2\pi t}{t_f}\right)-\pi-\frac{\varphi}{2}+\phi_{gb},
\end{equation}
and for the $\lambda(t)$ phase,
\begin{equation}
    \lambda(t)=\arctan\left[\left(1+\frac{\varphi}{2\pi}\right)\sin\left(\frac{2\pi t}{t_f}\right)\right].
\end{equation}
while the amplitude reads,

\begin{equation}
    F(t) = \frac{2\pi}{\mu t_f} \sqrt{1 + \left(1+\frac{\varphi}{2\pi}\right)^2\sin^2 \left( \frac{2\pi t}{t_f} \right)}.
\end{equation}
We have a family of time-dependent amplitude and frequency control fields which depends on the constant phase $\phi_{gb}$, all producing the same phase gate. Note in particular that in order to produce the $i\sigma_z$ gate, we simply set $\varphi=-\pi$ in the above formulas.




\subsection{Hadamard gate}
Consider the $SU(2)$ version of the Hadamard gate $iH_d$,
\begin{equation}
    U_{\rm gate} = 
     \frac{i}{\sqrt{2}}\begin{pmatrix} 1 & 1 \\ 1 & -1 \end{pmatrix} = iH_d
\end{equation}
We have for this gate $a_g=1/\sqrt{2}$ and $\phi_{ga}=\phi_{gb}=\pi/2$. Thus, we set $\theta=\pi/2$ which leads to $\chi= \frac{\pi^2}{t_f(\pi - 2)}$ and phase $\phi(t)$,
\begin{equation}
    \phi(t)= -\frac{2\pi }{(\pi-2)} \sin\left(\frac{\pi t}{2t_f}\right)+\frac{\pi^2}{\pi-2},
\end{equation}
while we get for the $\lambda(t)$ phase and $F(t)$,

\begin{equation}
    \lambda(t)=\arctan\left[\frac{2\pi}{\pi-2}\sin\left(\frac{\pi t}{2t_f}\right)\right],
\end{equation}

\begin{equation}
    F(t) = \frac{\pi}{2\mu t_f} \sqrt{1 + \frac{4\pi^2}{(\pi -2)^2} \sin \left( \frac{\pi t}{2t_f} \right)}.
\end{equation}
Note that since for the Hadammard gate the phases $\phi_{ga}$ and $\phi_{gb}$ have definite values, there is only a single control field to produce this gate for the cosine form of $a(t)$.

\subsection{Comparison with the full dynamics}

We illustrate the quality of the gates produced with the derived fields evolving the system under the complete Hamiltonian of \eqref{Ham} (without the RWA) and comparing the resulting dynamics of the evolution operator with the expected one from the RWA approach. We also calculate the gate fidelity as

\begin{equation}
   \mathcal{F}= \frac{1}{4} \left| \operatorname{Tr}\left\{U_{\rm gate}^\dagger U(t_f,0) \right\}\right|^2.
\end{equation}

\begin{figure}[H] 
    \centering 
    \includegraphics[width=1.0\textwidth]{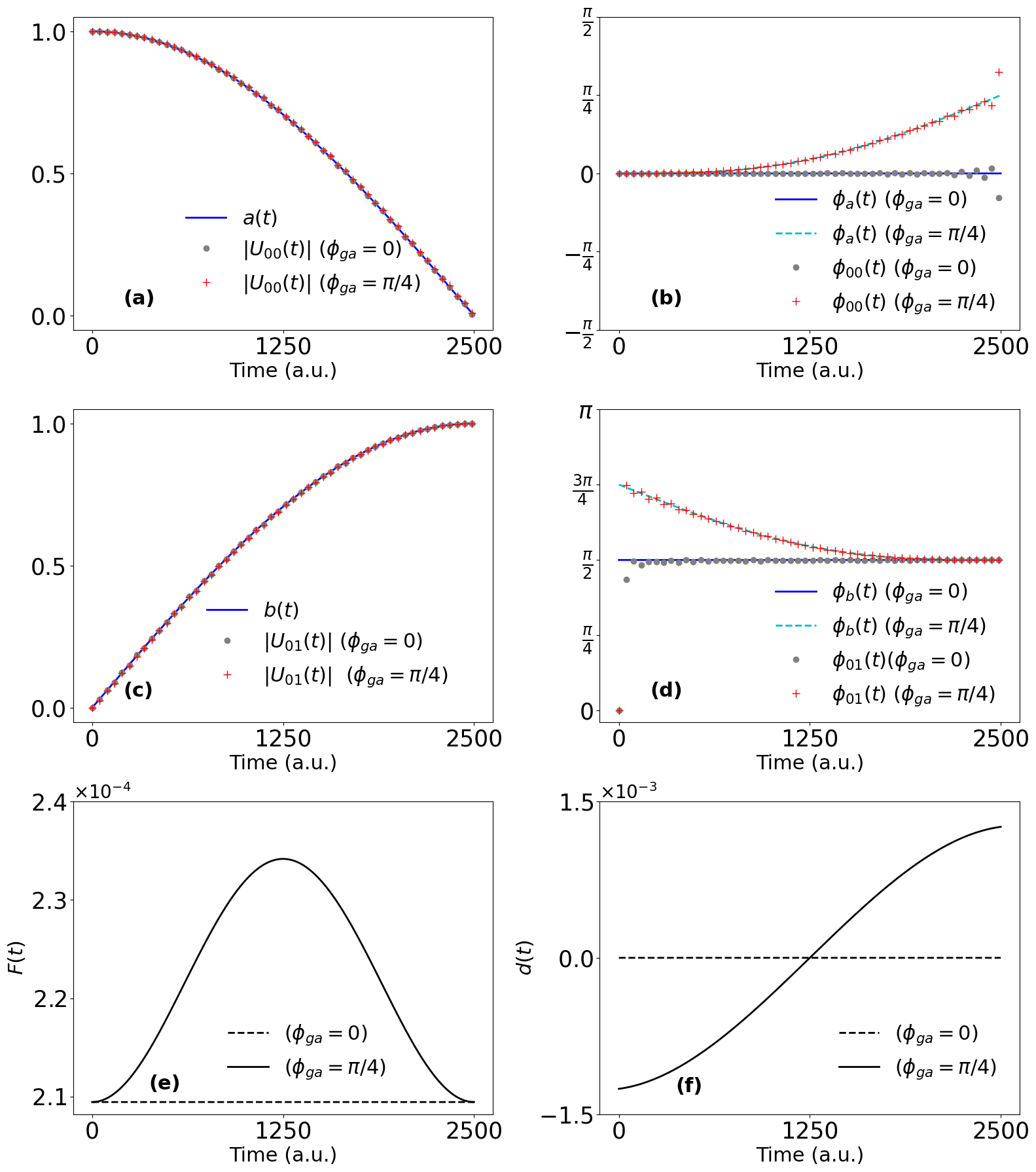} 
    \caption{Comparison between the chosen dynamical functions for the evolution operator and the numerical simulation for the implementation of the $i\sigma_x$ gate. The results are shown for $\phi_{ga} = 0$ and $\phi_{ga} = \pi/4$: (a) absolute value of the matrix element $U_{00}(t)$ and the function $a(t)$, (b) corresponding phase $\phi_{00}(t)$ and the function $\phi_a(t)$, (c) absolute value of the matrix element $U_{01}(t)$ and the function $b(t)$, (d) corresponding phase $\phi_{01}(t)$ and the function $\phi_b(t)$, (e) control field amplitude $F(t)$, and (f) field detuning $\Delta(t)$. The calculated fidelity is approximately $0.999990$ (for $\phi_{ga} = 0$) and $0.999968$ (for $\phi_{ga} = \pi/4$).}
    \label{fig:Xgate}
\end{figure}

Figure~\ref{fig:Xgate} shows the dynamics of the evolution operator $U(t,0)$ calculated for two fields producing the gate $i\sigma_x$, one with $\phi_{ga}=0$ ($\pi$-pulse) and the other with $\phi_{ga} = \pi/4$. We obtain very high fidelity with these fields, $0.999990$ for $\phi_{ga}=0$ and $0.999968$ for $\phi_{ga} = \pi/4$. Panels (a) and (b) refer to the first element of the $U(t,0)$ matrix, which we termed $U_{00}(t)$. Panel (a) shows the absolute value of $U_{00}(t)$ compared with the chosen dynamical function $a(t)$, while panel (b) shows the corresponding phase of $U_{00}(t)$, denoted by $\phi_{00}(t)$, compared with the chosen dynamical function $\phi_a(t)$. Panels (c) and (d) correspond, respectively, to the absolute value and phase of the second element of the evolution operator matrix $U_{01}(t)$. We can observe the good agreement with the chosen dynamical functions, the only unimportant discrepancy being when the phases are undefined. We note that for $\chi\neq0$ the phases are no longer constants. Panels (e) and (f) show that to modify the phases, it is required lager amplitude and also a chirping of the frequency. 

\begin{figure}[H] 
    \centering 
    \includegraphics[width=1.0\textwidth]{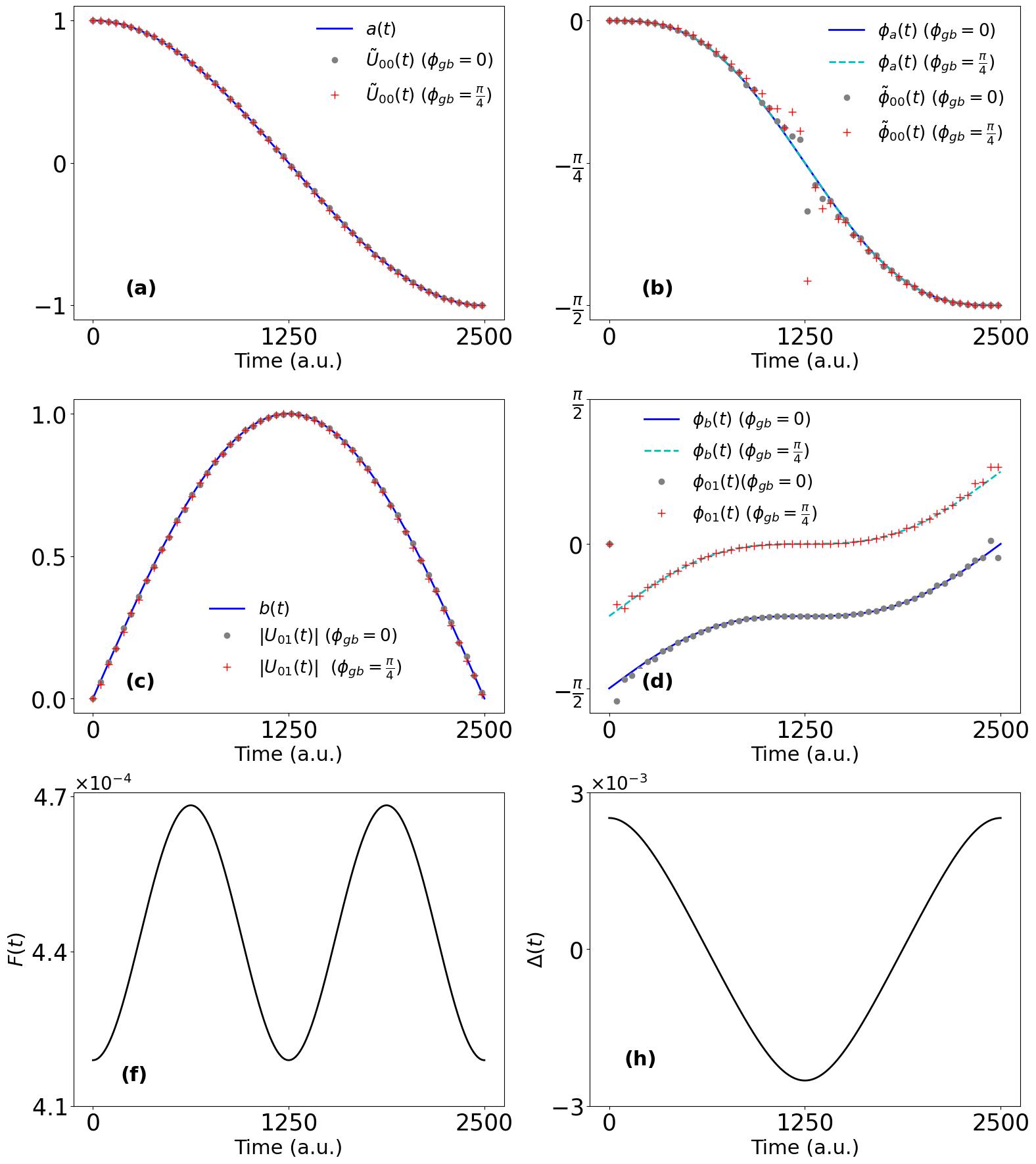} 
    \caption{Comparison between the chosen dynamical functions for the evolution operator and the numerical simulation for the implementation of the $i\sigma_z$ gate. The results are shown for $\phi_{gb} = 0$ and $\phi_{gb} = \pi/4$: (a) matrix element $\tilde{U}_{00}(t)$ and the function $a(t)$, (b) corresponding phase $\tilde{\phi}_{00}(t)$ and the function $\phi_a(t)$, (c) absolute value of the matrix element $U_{01}(t)$ and the function $b(t)$, (d) corresponding phase $\phi_{01}(t)$ and the function $\phi_b(t)$, (e) control field amplitude $F(t)$, and (f) field detuning $\Delta(t)$. Fidelidade: $0.9999086$ ($\phi_{gb} = 0$) e $0.9999528$ ($\phi_{gb} = \pi/4$). The calculated fidelity is approximately $0.9999086$ (for $\phi_{gb} = 0$) and $0.99995$ (for $\phi_{gb} = \pi/4$).}
    \label{fig:Zgate}
\end{figure}

In Fig.~\ref{fig:Zgate}, we consider the dynamics $i\sigma_z$ gate for two different values of $\phi_{gb}$. Again,  we observe good agreement with the chosen dynamical functions,  with very high fidelity, $0.9999086$ for $\phi_{gb}=0$ and $0.99995$ for $\phi_{gb} = \pi/4$. For the sake of comparison, in panel (b), we have shifted the numerically calculated phase $\phi_{00}(t)$ by $\pi$ for $t\geq t_f/2$ ( we indicate this modification with a \textasciitilde superscript). Note that the discontinuity of the first derivative of $|a(t)|$ at $t=t_f/2$ is compensated by a jump in the phase $\phi_a(t)$. In this panel, the numerically calculated phase become scattered at $t=t_f/2$ since $a(t_f/2)=0$, which leads to an indeterminate phase. Panel (d) shows that the different value of the constant phase $\xi$ only leads to a shift of $\phi_b(t)$. Panels (e) and (f) show, respectively, the necessary amplitude modulation and non-monotonic detuning to produce the gate.

\begin{figure}[H] 
    \centering 
    \includegraphics[width=1.0\textwidth]{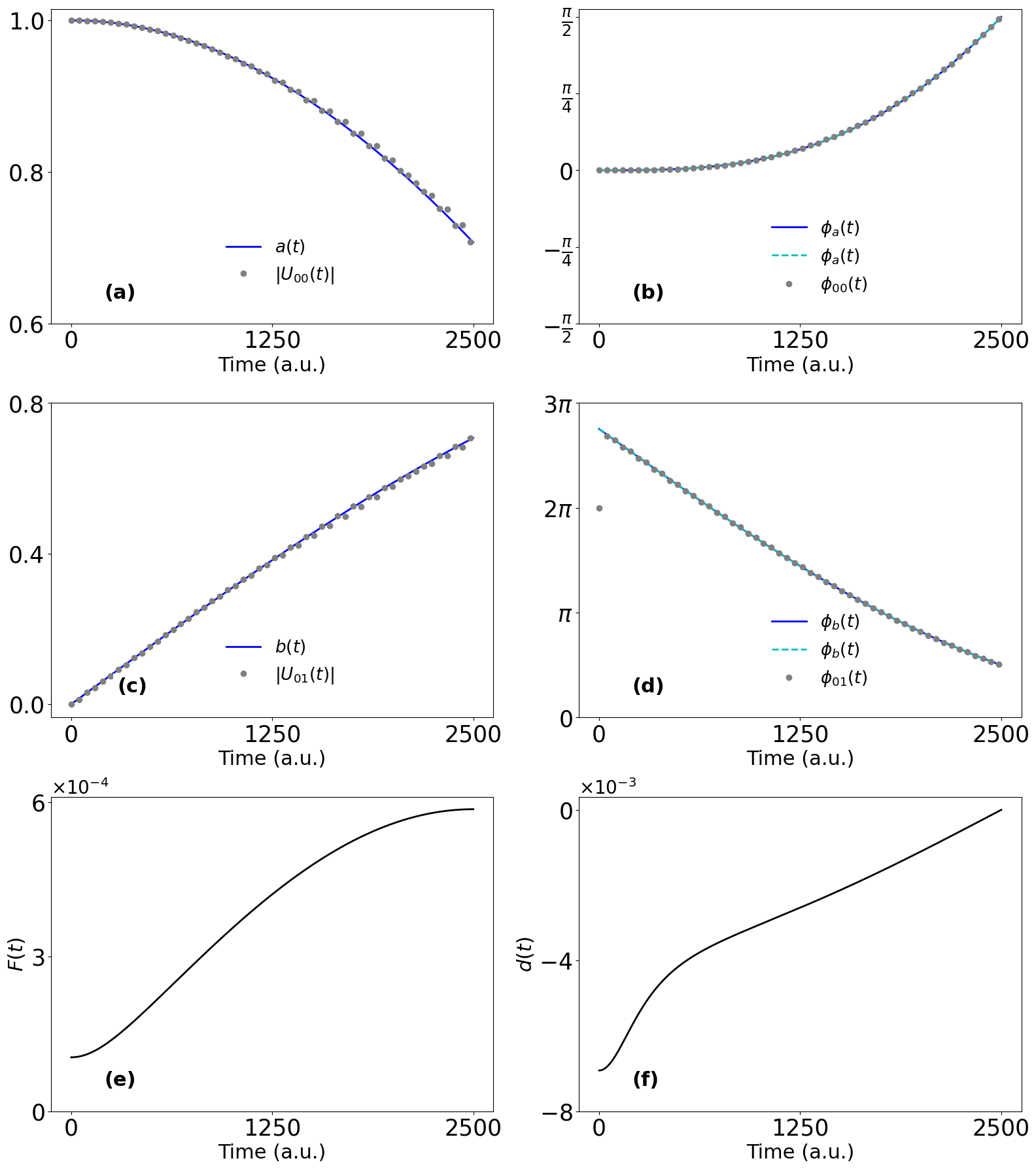} 
    \caption{Comparison between the chosen dynamical functions for the evolution operator and the numerical simulation for the implementation of the Hadamard gate $iH_d$: (a) absolute value of the matrix element $U_{00}(t)$ and the function $a(t)$, (b) corresponding phase $\phi_{00}(t)$ and the function $\phi_a(t)$, (c) absolute value of the matrix element $U_{01}(t)$ and the function $b(t)$, (d) corresponding phase $\phi_{01}(t)$ and the function $\phi_b(t)$, (e) control field amplitude $F(t)$, and (f) field detuning $\Delta(t)$. The  calculated fidelity is approximately $0.99986$}
    \label{fig:Hadgate}
\end{figure}

Figure~\ref{fig:Hadgate} shows the results for the Hadamard gate $iH_d$. The agreement with the numerical calculation of the evolution operator is noticeable in panels (a) to (d). In panels (e), we see that the amplitude increases monotonically, while panel (f) shows that the detuning increases approximately linear for $t>t_f/3$. The calculated fidelity is $0.99986$.

\pagebreak

\pagebreak

\pagebreak

\section{Optimization of the control field}

Free parameters of the dynamical functions can be used to minimize a given cost function. We illustrate the optimization of the parameters considering the the pulse fluency as a cost functional to be minimized,
\begin{equation}
    J\{\bm{\alpha}\} = \int^{t_f}_0 E(\bm{\alpha},t)^2 dt,
\end{equation}
where $\bm{\alpha}$ stands for a set of free parameters that can be tuned to minimize the cost functional. As we have seen above, for the $i\sigma_x$ gate, the parameter $\phi_{ga}$ is a free parameter, independent of the choice of $a(t)$, since $\phi_{ga}$ is undefined for the gate. In this situation, we term $\phi_{ga}$ a naturally free parameter for the $i\sigma_x$ gate, or $\bm{\alpha}=\phi_{ga}$. For the cosine solution, $\phi_{ga}=0$ yields the square $\pi$-pulse, while different values of this phase implies in pulses with modulated amplitudes and frequency. For the phase gates, $\phi_{gb}$ is a naturally free parameter, since its values is undefined, $\bm{\alpha}=\phi_{gb}$. However, with limited effect on the laser fluency, since it contributes only with a constant phase for the pulse.

The cosine \textit{anzatz} for $a(t)$ does not furnish much flexibility to the control pulse, having only the naturally free parameter. But we can chose other functions for $a(t)$ with additional free parameters. As an example, we consider an alternative form of $a(t)$ for the $i\sigma_z$ gate. For this gate, we specify  $a_{g}=1$ and $\phi_{ga}=\pi/2$. Then a convenient form for $a(t)$ is the flipped bell-shaped function,  
\begin{equation}
    a(t) = \sqrt{1 - \delta^2 \left(\frac{t}{t_f}\right)^2 \left(1 - \frac{t}{t_f}\right)^2},\label{eq:a_raiz}
\end{equation}
which satisfies $a(0)=1$ and $a(t_f)=1$. The parameter $\delta$ is such that $0<\delta<4$ and controls the depth of the curve at $t=t_f/2$. Note that $\delta$ is a free parameter, in addition to the naturally free parameter $\phi_{gb}$, thus we have $\bm{\alpha}=(\delta,\phi_{gb})$. The corresponding form for $b(t)$ is,
\begin{equation}
    b(t) = \delta \frac{t}{t_f} \left(1 - \frac{t}{t_f} \right).
\end{equation}

As before, we chose $\dot{\phi}_a(t)$ in such a way as to eliminate the time dependent part of the denominator of \eqref{eq:cond_phi},

\begin{equation}
    \dot{\phi}_a(t)=\alpha\delta^2 \left(\frac{t}{t_f}\right)^2\left(1-\frac{t}{t_f}\right)^2,
\end{equation}
with $\alpha$ being a constant. Integrating and applying the boundary conditions we obtain $\alpha=15\pi/\delta^2t_f$ and,
\begin{equation}
   \phi_a(t)
= 5\pi\left(\frac{t}{ t_f}\right)^3- \frac{15\pi}{2} \left(\frac{t}{ t_f}\right)^4+3\pi \left(\frac{t}{ t_f}\right)^5.
\end{equation}
From \eqref{eq:cond_phi}, we have

\begin{equation}
   \dot{\phi}_b(t)= \frac{15\pi}{t_f\delta^2}\left[\delta^2\left(\frac{t}{t_f}\right)^2 \left(1 - \frac{t}{t_f}\right)^2-1\right].
\end{equation}
Integrating,

\begin{equation}
\phi_b(t)=\phi_a(t) +\frac{15\pi}{\delta^2}\left(1-\frac{t}{t_f}\right)
-\frac{\pi}{2}+\phi_{gb}.
\end{equation}

We obtain for the amplitude,
\begin{align}
    & F(t)= \frac{2}{\mu t_f} \sgn\!\left(1-2\frac{t}{t_f}\right)\times \nonumber \\
& \sqrt{\frac{\delta^2\left(1-2\frac{t}{t_f}\right)^2}{\left[1-\delta^2 \left(\frac{t}{t_f}\right)^2 \left(1-\frac{t}{t_f}\right)^2\right]}+\left(\frac{15\pi}{\delta}\right)^2 \left( 1 - \delta^2 \left( \frac{t}{t_f} \right)^2  \left(1- \frac{t}{t_f} \right)^2 \right)\left(\frac{t}{t_f}\right)^2\left(1-\frac{t}{t_f}\right)^2},
\end{align}
while the phase $\lambda(t)$ reads,

\begin{equation}
    \lambda(t) = \arctan\left\{ \frac{15\pi}{\delta^2} \left[ \delta^2 \left(\frac{t}{t_f}\right)^2 \left(1 - \frac{t}{t_f}\right)^2 -1\right] \frac{\frac{t}{t_f}\left(1 - \frac{t}{t_f}\right)}{1 - 2\frac{t}{t_f}} \right\}.
\end{equation}

\begin{figure}[H] 
    \centering 
    \includegraphics[width=1.0\textwidth]{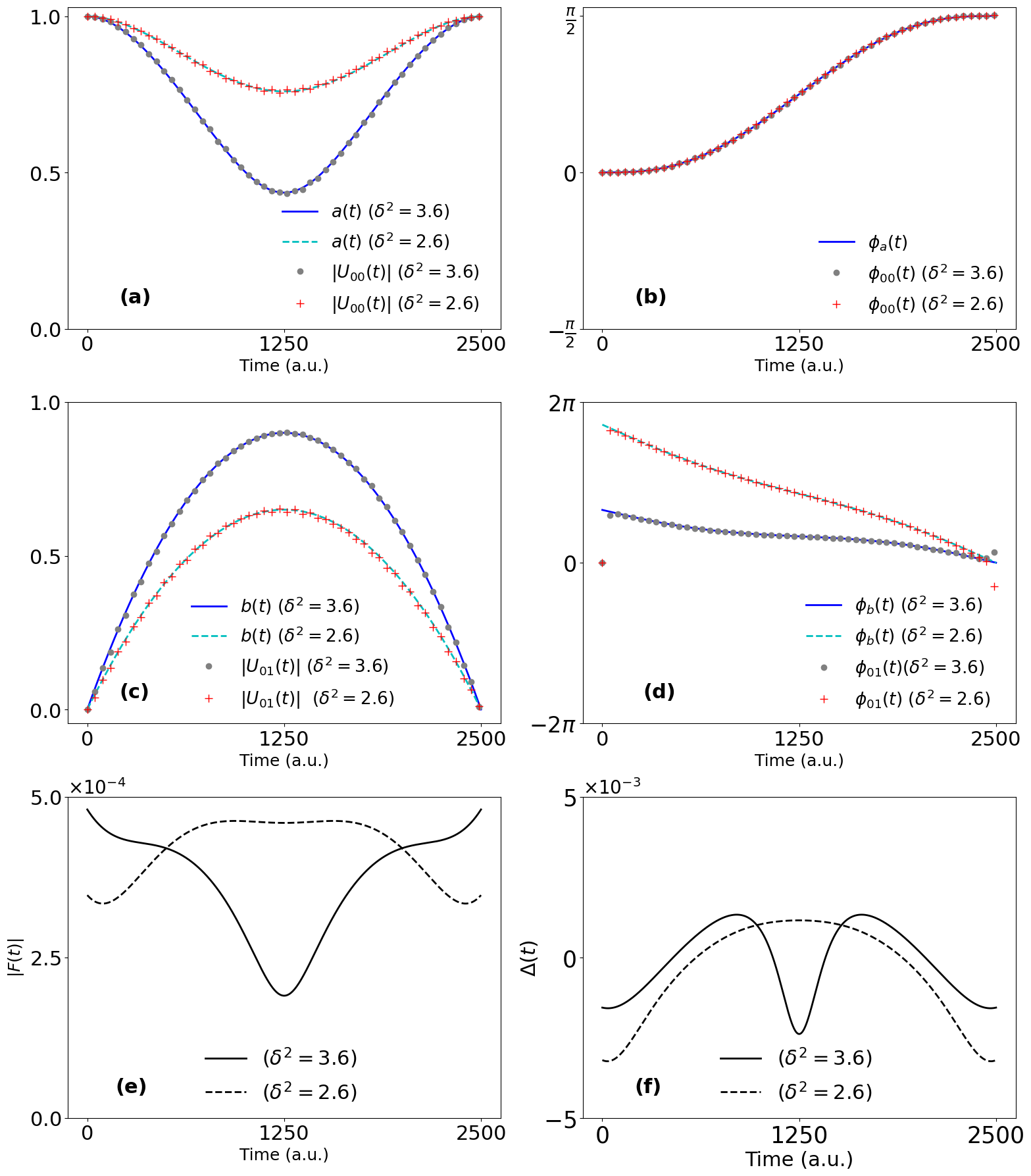} 
    \caption{Comparison between the chosen dynamical functions for the evolution operator and the numerical simulation for the implementation of the $i\sigma_z$ gate using the square-root solution. The results are shown for $\delta^2 = 3.6$ and $\delta^2 = 2.6$: (a) absolute value of the matrix element $U_{00}(t)$ and the function $a(t)$, (b) corresponding phase $\phi_{00}(t)$ and the function $\phi_a(t)$, (c) absolute value of the matrix element $U_{01}(t)$ and the function $b(t)$, (d) corresponding phase $\phi_{01}(t)$ and the function $\phi_b(t)$, (e) control field amplitude $F(t)$, and (f) field detuning $\Delta(t)$. Fidelity: $0.9998284$ ($\delta^2 = 3.6$), $0.9998626$ ($\delta^2 = 2.6$)}
    \label{fig:SigZ_raiz}
\end{figure}

Figure~\ref{fig:SigZ_raiz} shows the dynamics of the evolution operator corresponding to the square-root solution \eqref{eq:a_raiz}. Panels (a) to (d), shows the good concordance of the dynamics functions $a(t)$, $b(t)$, $\phi_{ga}$ $\phi_{gb}$ with the numerical ones for two different values of $\delta$ ($\delta^2 = 3.6$ and $\delta^2 = 2.6$). In panels (e) and (f), we can notice the great impact of changing $\delta$ in the amplitude and detuning profiles.

\begin{figure}[H] 
    \centering 
    \includegraphics[width=0.55\textwidth]{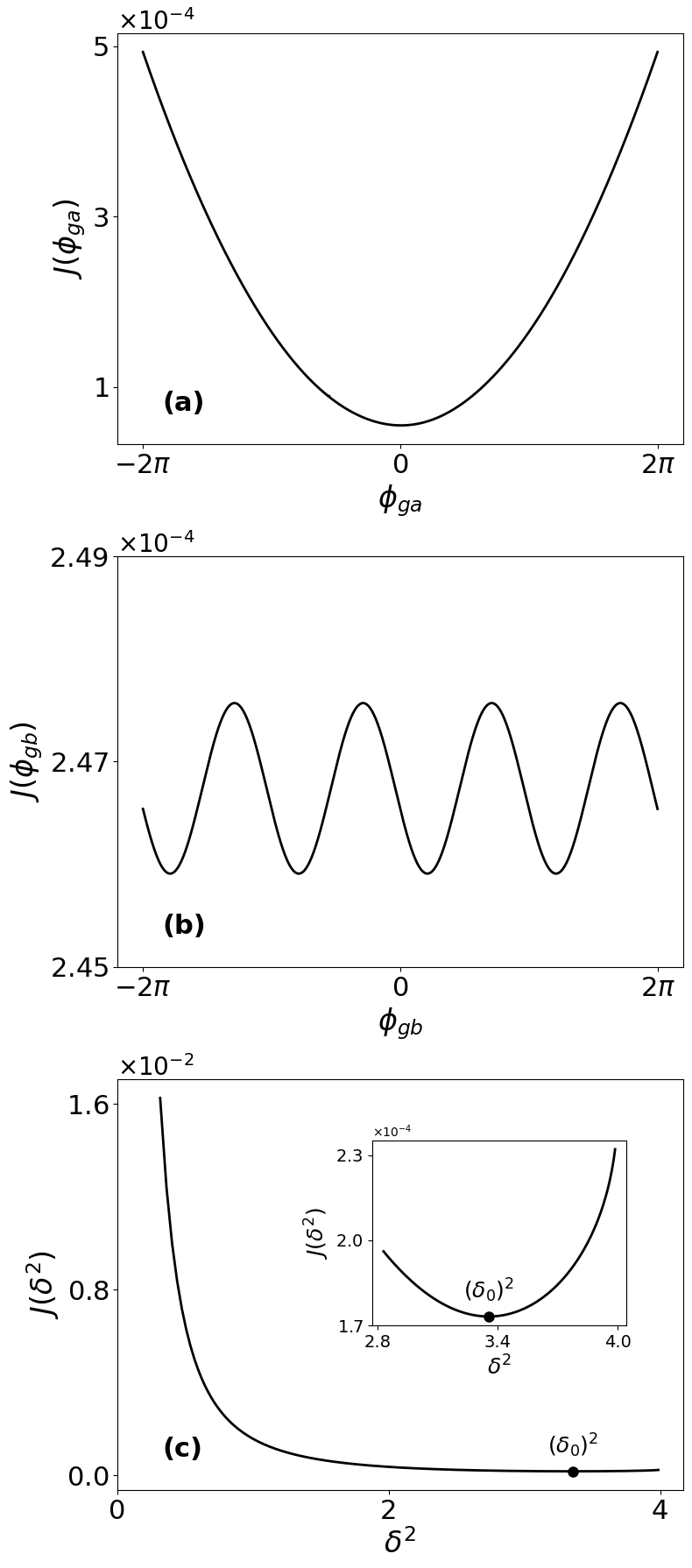} 
    \caption{Pulse fluency $J$ as a function of the free parameters for the optimization of the control field: (a) $i\sigma_x$ gate implemented with the cosine solution as a function of the free parameter $\phi_{ga}$. (b) $i\sigma_z$ gate implemented with the cosine solution as a function of the free parameter $\phi_{gb}$. (c) $i\sigma_z$ gate implemented with the square-root solution as a function of the parameter $\delta^2$ and the inset highlights the minimum value reached at $(\delta_0)^2 \approx 3.36$.}
    \label{fig:minJ}
\end{figure}

Figure~\ref{fig:minJ} shows the values of the fluency as a function of the free parameters. Panel (a) shows the fluency of the $i\sigma_x$ gate as a function of $\phi_{ga}$. As could be expected, the square $\pi$-pulse with $\phi_{ga}=0$ is the optimal choice, yielding the lowest fluency. Panel (b) shows the fluency of the $i\sigma_z$ gate for the cosine solution as a function of $\phi_{gb}$. We note that the $\phi_{gb}$ has small impact on the fluency since it just amounts to different constant phase to the control field. Panel (c) considers the $i\sigma_z$ gate as a function of $\delta$ for the square-root solution \eqref{eq:a_raiz}. By contrast to $\phi_{gb}$, changing the parameter $\delta$ has a considerable effect on the fluency. We found the optimal $\delta$ value being $\delta_0^2\approx3.36$, corresponding to a fluency of $1.7\times10^{-4}$, which can be compared to the minimum fluency obtained with the cosine solution of $2.46\times10^{-4}$.

\begin{figure}[H] 
    \centering 
    \includegraphics[width=0.8\textwidth]{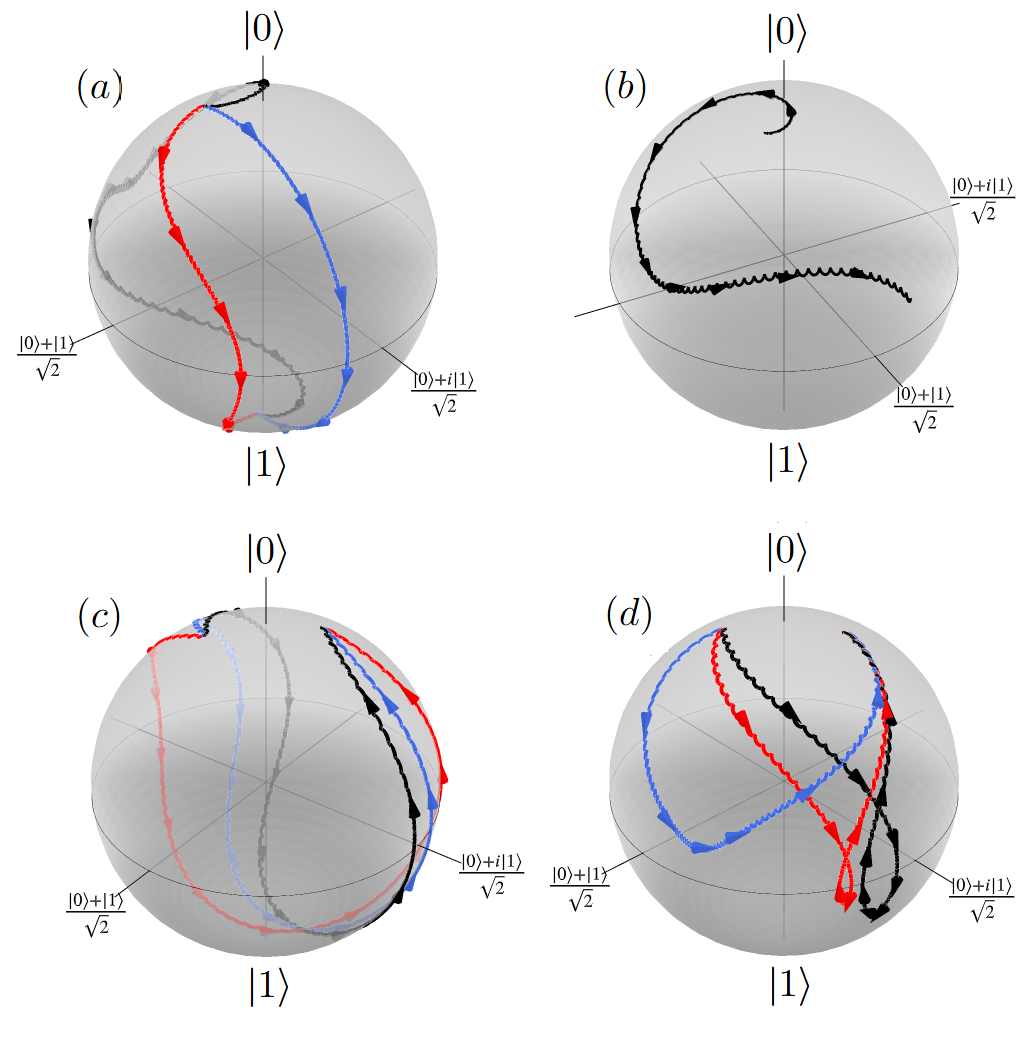} 
    \caption{Bloch sphere trajectories obtained from the numerical simulations for an initial state with ground-state population $\rho_{gg}(0)=0.8$, coherence magnitude $|\rho_{ge}(0)|=0.8$, and relative phase $\phi(0)=\pi/4$. All trajectories share the same initial condition. (a) $i\sigma_x$ gate for three values of the free parameter: $\phi_{ga}=0$ (blue), $\phi_{ga}=\pi/2$ (red), and $\phi_{ga}=2\pi$ (black). (b) $iH_d$ gate. (c) $i\sigma_z$ gate implemented with the cosine solution for $\phi_{gb}=0$ (blue), $\phi_{gb}=\pi/6$ (red), and $\phi_{gb}=\pi/4$ (black). (d) $i\sigma_z$ gate implemented with the square-root solution for $\delta^2=3$ (blue), $\delta^2=\delta_0^2$ (red), and $\delta^2=3.5$ (black).}
    \label{fig:esferas}
\end{figure}

Figure~\ref{fig:esferas} shows the Bloch sphere trajectories obtained from the numerical simulations for a fixed initial state characterized by a ground-state population of $0.8$, coherence magnitude of $0.8$, and coherence phase of $\pi/4$. For the $i\sigma_x$ gate, shown in panel (a), the trajectory associated with $\phi_{ga}=0$ (blue) follows the most direct evolution between the initial and final states, avoiding the additional loops observed for larger values of $\phi_{ga}$. This behavior is consistent with the optimization analysis presented in Fig.~\ref{fig:minJ}(a), where $\phi_{ga}=0$ minimizes the pulse fluency and corresponds to the square $\pi$-pulse.
In contrast, increasing $\phi_{ga}$ forces the state to undergo additional rotations on the Bloch sphere before reaching the same final state, without changing the implemented logical operation. Panels (c) and (d) compare two different analytical solutions that realize the same $i\sigma_z$ gate. For the cosine solution, changing the naturally free parameter $\phi_{gb}$ modifies the trajectory, although all three evolutions remain similarly extended over the Bloch sphere. In contrast, the square-root solution provides greater flexibility through the additional free parameter $\delta^2$. As a consequence, the trajectories become more direct, involving fewer unnecessary rotations, with this effect being particularly evident for $\delta^2=\delta_0^2$, which corresponds to the minimum pulse fluency obtained in Fig.~\ref{fig:minJ}(c). These results indicate that optimizing the control field not only reduces the energetic cost of the pulse but also leads to a more efficient evolution of the quantum state on the Bloch sphere.

\pagebreak

\section{conclusion}

In this work, we provide a general framework leading to analytical formulas for linearly-polarized pulses that produce any desired single-qubit quantum gate within the RWA. In our formulation, we express the evolution operator in terms of dynamical functions that can be prescribed \textit{a priori}, $a(t)$ and $\phi_a(t)$. We then invert the equations of motion, leaving the control pulse in term of these dynamical functions. As we have shown, a key procedure for obtaining simple analytical solutions is to chose the derivative of dynamical function $\phi_a(t)$ proportional to $1-a(t)^2$. Also the dynamical function $a(t)^2$ is chosen sufficiently simple to allow for analytical integration. We have verified the success of the approach with some specific choices for the dynamical functions. Furthermore, it is possible to chose the function $a(t)$ depending on free parameters that can be adjusted \textit{a posteriori}. Therefore, besides of matching the initial and final conditions for the gate implementation, there are still free parameters that can be used to minimize a given cost functional.


\end{document}